\documentclass[fleqn,usenatbib]{mnras}

\usepackage[T1]{fontenc}

\DeclareRobustCommand{\VAN}[3]{#2}
\let\VANthebibliography\thebibliography
\def\thebibliography{\DeclareRobustCommand{\VAN}[3]{##3}\VANthebibliography}

\usepackage{graphicx}	
\usepackage{amsmath}	
\usepackage{amssymb}	

\usepackage{newtxtext,newtxmath} 

\usepackage{dsfont}
\usepackage[tracking=true]{microtype}
\SetTracking[spacing={-50*,0*,50*}]{encoding=OT1}{-300}
\newcommand{\spar}{S_{\textls{//}}}
\newcommand{\sper}{S_{\bot}}
\newcommand{\vspar}{\vec{S}_{\textls{//}}}
\newcommand{\vsper}{\vec{S}_{\bot}}

\title[Formalism for the analysis of alignments]{A statistical formalism for alignment analysis}

\author[F. Dávila-Kurbán et al.]{
F. Dávila-Kurbán,$^{1,2,3,4}$
M. Lares,$^{1,2,3}$\thanks{E-mail: marcelo.lares@unc.edu.ar}
D. Garcia Lambas$^{1,2,3}$
\\
$^{1}$Instituto de Astronom\'ia Te\'orica y Experimental (IATE,
        CONICET/UNC), Córdoba, Argentina\\
$^{2}$ Observatorio Astron\'omico C\'ordoba, Argentina\\
$^{3}$ Consejo de Investigaciones Científicas y Técnicas (CONICET),
Argentina \\
$^{4}$ Facultad de Matemética, Astronomía y Física, Universidad
Nacional de Córdoba, Argentina
}

\date{Accepted XXX. Received YYY; in original form ZZZ}

\pubyear{2021}

\begin{document}
\label{firstpage}
\pagerange{\pageref{firstpage}--\pageref{lastpage}}
\maketitle

\begin{abstract}
The detection of anisotropies with respect to a given direction in a vector field is a common problem in astronomy.
Several methods have been proposed that rely on the distribution of the acute angles between the data and a reference direction.
Different approaches use Monte Carlo methods to quantify the statistical significance of a signal, 
although often lacking an analytical framework.
Here we present two methods to detect and quantify alignment signals
and test their statistical robustness.
The first method considers the deviance of the relative fraction of vector components in the plane perpendicular to a reference direction with respect to an isotropic distribution.
We also derive the statistical properties and stability of the resulting estimator,
and therefore does not rely on Monte Carlo simulations to assess its statistical significance.
The second method is based on a fit over the residuals of the empirical cumulative distribution function with respect to that expected for a uniform distribution,
using a small set of harmonic orthogonal functions, which does not rely on any binning scheme.
We compare these methods with others commonly used in the literature,
using Monte Carlo simulations,
finding that the proposed statistics
allow the detection of alignment signals with greater significance.
\end{abstract}


\section{Introduction}

Shapes and orientations of galaxies and the large scale structures in which they are embedded may have significant coherence given the effects of accretion and mergers as well as tidal, stripping and other combined actions.
As a consequence of these processes, the statistical properties of the galaxy orientations with respect to the
cosmic web structures may differ from those expected for randomly oriented galaxies (e.g, see \citealt{Mo2010}).
Thus, studies of intrinsic alignment signals allow to explore the links
between the joint evolution of galaxies
and their surrounding structures \citep[e.g.][and
references therein]{Panko2013}.
Taking into account these facts, the analysis of the orientations of galaxies in the context of both the local environment and the large--scale structures may be crucial to test scenarios of galaxy formation and evolution, in particular for
theoretical predictions of the angular momentum
of galaxies \citep[e.g.,][]{Peebles1969}.
The alignment signals, however, are somewhat elusive, given the variety
of preferred directions that arise from the actual distribution of surrounding structures and
the fact that galaxy orientations with respect to any direction are
mainly random.
For these reasons a robust statistical method to detect and assess alignment
signals and their significance is a key tool in studies of alignments
between galaxies and the large-scale distribution of
structures.


In the case of spiral galaxies, the spatial distribution of stars in a
disc defines a preferred plane
whose normal is oriented roughly onto the rotation axis.
The tidal field exerted by regions characterized by structures
such as clusters, filaments or voids, are present during a considerable extent
of galaxy evolution, and could produce observable features in their
original spin vector. The fact that galaxies rotate are indicative of the physical conditions under which they formed, and the rotation itself is certainly an important test of any theory for the origin of the galaxies \citep{Peebles1969}. Galaxy angular momentum is widely believed to arise from gravitational torque due to misalignment of the gravitational shear tensor and the inertia tensor in early formation stages \citep{Doroshkevich1970, White1984}. Thus, the galaxy spin field holds information about the gravitational shear field and can be used, for example, for a statistical reconstruction thereof \citep{Lee2000}.
Furthermore, it is commonly assumed that, during early stages of formation, baryonic and dark matter shared a similar evolution and likely gained the same specific angular momentum prior to the formation of the disc \citep[e.g.,][]{FallEfsthathiou1980}. The study of alignment dark matter haloes (hereafter DM haloes) was possible, and became a popular subject, after N-body simulations had enough resolution to perform studies of this nature \citep[e.g.][and references therein]{Cuesta2008,Libeskind2013,Forero-Romero2014,Joachimi2015,Kiessling2015}.


The methods of alignment detection and the results obtained are diverse. For example, \citet{Forero-Romero2014} studied alignment of shape, angular momentum, and peculiar velocity of DM haloes with respect to the cosmic web, as described by using the tidal field or velocity shear, employing the Bolshoi simulation \citep{Riebe2011}.
%
They quantify the alignments by measuring the fraction of haloes that is preferentially aligned with one of the eigenvectors in the local definition of the cosmic web, and with the average value of the angle between an eigenvector and the vector of interest.
%
They found the strongest alignment for halo shapes with filaments and walls defined by the tidal field, but when defined by velocity shear they found anti--alignment with massive haloes. 
%
For the angular momentum, they only found a weaker signal for the most massive haloes to be anti--aligned with filaments, and being aligned along the sheets of the velocity shear. There is a discrepancy with previous works \citep{AragonCalvo2007, Hahn2007, AragonCalvoYang2014} which indeed detect alignments for less massive haloes. \citet{Forero-Romero2014} argues that this might be due to high sensitivity of the alignment signal to the small-scale cosmic web description.
%
Additionally, they find peculiar velocities to be preferentially parallel to walls and filaments.
%
These results indicate that the alignment properties of DM haloes can depend on the physical definition of the cosmic web, with tidal field versus velocity shear approach yielding complementary information.

With the greater computing power of recent years, the study of alignments of galaxies in simulation has been possible. \citet{Codis2018} and \citet{Kraljic2019}, for example, study the distribution of angles measured between the spin of galaxies and haloes and the different elements of the surrounding cosmic web in the Horizon--AGN and SIMBA simulations, respectively. Their results agree on the spin of low-mass galaxies being more likely to lie within the plane of sheets while massive galaxies preferentially having a spin perpendicular to the sheets.



%
The search for galactic alignment has been analyzed also in observations in the context of structures that, to a reasonable extent, can be described with spherical symmetry, like clusters of galaxies or voids.
The observational aspect of this topic of study has its own difficulties to face, mainly the small sample sizes and line-of-sight projection effects.
Earlier works focused on the orientation of galaxies with respect to the Local Supercluster and other clusters such as Virgo and Coma (e.g. \citealt{KashikawaOkamura1992, Godlowski1993, Godlowski1994, Hu1995, Hu1997, Yuan1997, Hu1998, GodlowskiOstrowski1999}) relied on the "position angle (PA) -- inclination method" \citep{JaanisteSaar1978, FlinGodlowski1986}. In this method, the measured PAs of galaxies (usually on photographic plates) are converted into 3-dimensional vectors using inclination angles obtained from the measured projected minor-to-major axial ratios, b/a. The distribution of these vectors could then be compared with a null-hypothesis, e.g. isotropic spatial distribution, and thus assess whether the data is isotropic or anisotropic by comparison.
However, the shape of these isotropic distributions can be significantly affected by selection criteria \citep{Aryal2000}. These effects can be large when the sample is selected from incomplete datasets (e.g. a limited portion of the sky) and lead to artificial structures in the data. 
Therefore, a statistically robust method that reliably describes not only the data, but the comparison sample as well, is of crucial importance in these analyses in order to conclude in favor of either isotropy or anisotropy in the data
.




Other observational studies employ similar methods that also rely on binning statistics such as the normalized pair count, P(cos$\theta$) in bins of the measured angle $\theta$ between the subject of interest (e.g. satellite or central galaxies or otherwise) and a preferred direction determined by some other structure such as cluster centers or elements of the cosmic web \citep[e.g.][]{Brainerd2005, Yang2006, Zhang2015}. The significance of these statistics is usually assessed by comparison with a large number of Monte Carlo simulations. 


\citet[][hereafter V12]{Varela2012} performed a rigorous assessment of an analytical model for the distribution of $\theta$, and its behaviour in the isotropic case for the estimation of the statistical significance, based on previous works \citep[e.g.][]{BetancortRijoTrujillo2009, Brunino2007, Cuesta2008, Lee2004}.
This work tackled some discrepancies that emerged in previous observational studies of the alignment of galaxies around voids, namely \citet{Trujillo2006} and \citet{Slosar2009} (hereafter T06 and S09, respectively). T06 analyzed 201 face-on and edge-on galaxies using data from the SDSS-DR3 and the 2dFRGS \citep{Colless2001} and found a significant tendency of the spin of the galaxies to be in the direction perpendicular to the void radial direction. On the other hand, S09 using two samples of 578 and 258 galaxies from the SDSS-DR6 with similar selection criteria found no statistical evidence for departure from random orientations. 
V12, used the SDSS-DR7 and a statistical procedure robust enough to overcome the problem of the indeterminacy of the real inclination of galaxies computed from their apparent axial ratio, and assess the validity of the procedure with extensive Monte Carlo simulations.
%
They detect a statistically significant tendency of galaxies around large voids (with radii of over 15 h$^{-1}$Mpc) to have their angular momenta aligned with the radial direction of the voids.
%
%
This highlights the importance of, not only a bigger sample size, but the use of robust and reliable statistical methods to correctly assess the validity of the alignment signal detected.

In this paper we present two formal methods to analyze the alignments
of a sample of particles with respect to a center.
The first method consists on the definition of simple metrics from the
radial and tangential components of the vectors, 
while the second one relies on the parametrization of a residual function between the data obtained from the sample and from an isotropic distribution.
On either case, we do not assume any binning scheme.
Instead, we use all the information in the data and apply robust
estimations of the uncertainties in the alignments metrics.
By deriving the theoretical distribution of the parameters that measure alignment signal we can not only determine its statistical significance with accuracy, but we can do so without investing computational resources and time into the Monte Carlo simulations usually needed to estimate this.


The outline of this paper is as follows. In Section~\ref{S_Method} we develop the statistical formalism for the aforementioned two new methods for the study of vector alignments, and introduce the parameters with which to measure alignment signal. In Section~\ref{S_Comparison} we apply the methods to synthetic data corresponding to different scenarios of alignment to test how well the new parameters recover the alignment signal, and compare it to more traditional methods. Finally, Section~\ref{S_Summary} presents our main conclusions.

\begin{figure*}
\includegraphics[width=\textwidth]{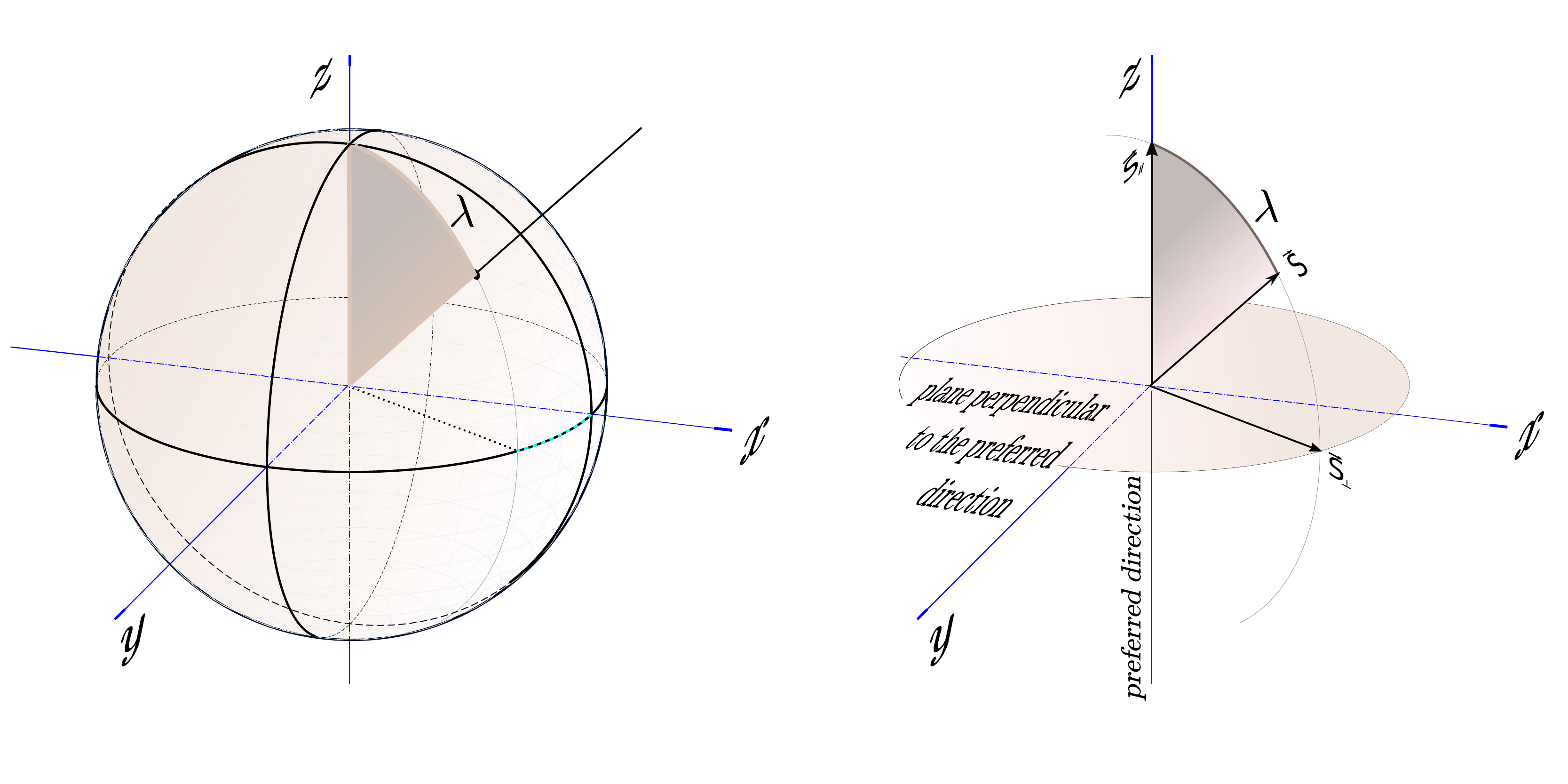}
\caption{Coordinate system used.  The $z$ axis is the radial outward direction
   of the void.  The angle $\theta$ is formed between the $z$ and the vector $\vec{S}$, 
   and takes values in the range $[0, \pi]$.}
   \label{F_sist_coords}
\end{figure*}

\section{Methods}
\label{S_Method}


The symmetry of structures such as filaments, haloes, The symmetry of structures such as filaments, haloes, clusters or voids, both in their geometry as in their dynamics, 

clusters or voids, both in their geometry as in their dynamics, 
allows the consideration of a preferential direction to analyze the orientation of galaxies. In the case of spherical symmetry, this is the radial direction.
The objective is to develop a statistical formalism to measure in a robust manner
the distribution of the orientation of galaxies and detect possible excesses with 
respect to a completely random distribution.
Given the problem of vector orientations with respect to a central point
we want to define a statistical parameter and obtain its distribution
in order to know the significance of a hypothesis test.

\subsection{Ratio of vector components}
\label{S_Beta}


Given a radial direction $\hat{z}$ of unit norm (see
Fig.~\ref{F_sist_coords}), perpendicular and
parallel components of vector $\vec{S}$ can be calculated as:

\begin{equation}
    \vspar = \vec{S} \cdot \hat{z}, \qquad \textrm{and} \qquad \vsper = \vec{S} - \vspar,
\end{equation}

\noindent
where $\vsper$ is the perpendicular component to the radial direction $\hat{z}$,
$\vspar$ is the parallel component to the radial direction and
$\vec{S}=\vsper+\vspar$.


The angle $\theta$ formed by the radial direction and the direction of the vector
$\vec{S}$ is related to the components:

\begin{equation}
    \sper = |\vec{S}|\;sin(\theta); \qquad \spar = |\vec{S}|\;cos(\theta).
\end{equation}


The distribution of this angle can be used to analyze alignments, and
given its relation to the components, the latter can also be used to determine
the orientations. To that effect we define:

\begin{equation}
    \mathcal{B} = \frac{\sper}{\spar} = \frac{S \sin(\theta)}{S \cos(\theta)} = \tan(\theta).
\end{equation}


The parameter $\mathcal{B}$ is also a measure of the orientation of the vector
$\vec{S}$. Note that the ranges of these two parameters are as follows:

$$
0 \le \theta \le \pi; \qquad -\infty \le \mathcal{B} \le \infty
$$


Using the symmetry of the problem, we can define the
parameters considering the acute angle between the directions
$\hat{z}$ and $\hat{S}$, as well as the norm of the component $\vspar$,

\begin{equation}
   \beta = |\mathcal{B}| = \frac{\sper}{|\spar|},\qquad \text{with} \qquad \lambda = min(\theta, \pi-\theta)
\label{E_beta}
\end{equation}

for which:

$$
0 \le \lambda \le \frac{\pi}{2}; \qquad 0 \le \beta \le \infty
$$


Vectors with $\beta>1$ have a preference in the
perpendicular direction and $\pi/4 < \lambda < \pi/2$, while
vectors with $\beta<1$ have a preference in the
radial direction and $0 < \lambda < \pi/4$.

To quantify the direction of $\vec{S}$ the following statistics could be considered:

\begin{itemize}
   \item the angle $\theta$ 
   \item the acute angle $\lambda$ 
   \item the ratio of perpendicular to parallel components,
      $\mathcal{B}$ (with $\theta$)
   \item the ratio of perpendicular to parallel components,
      $\beta$ (with $\lambda$)
\end{itemize}





We will explore in the following subsections the use of the angles or the ratios.
It is of utmost importance to establish the distributions of these parameters
for the case in which there is no alignment signal whatsoever. This way we determine the amplitude of the statistical fluctuations and establish a measurement of the signal in a data sample calculating its statistical significance.
To this end, we define the null hypothesis

$$H_0:\textrm{the distribution of vectors are random with spherical symmetry}$$

\noindent
i.e., there is no alignment signal whatsoever.
This hypothesis can also be used to generate control samples with Monte Carlo
procedures if need be.

Note that the regions of $\beta$ greater or lesser than 1 are different,
so it is expected that for a random distribution there would be more "perpendicular"
than "parallel" vectors (see below Fig.~\ref{F_rvs_pdf}, upper panel).
              


\subsection{Distribution of the ratio of vector components}
\label{S_BetaDist}

In this Section we derive the distribution of the test statistic $\beta$, defined as the ratio between the perpendicular and parallel vector components (Eq. \ref{E_beta}), under the null hypothesis.

The distribution of $\beta$ can be deducted from the change of random variables theorem \citep{gillespie_theorem_1983}, which, in its general form, can be enunciated as follows:

Let $\{X_i\}_{i=1}^n$ $n$ be a R.V. with known $f_{\vec{X}}(\vec{x})$,
and let $m$ R.V. $\vec{Y} = \varphi(\vec{x})$, where $\varphi=(\varphi_1,
\varphi_2, \ldots, \varphi_m)^t$ and $\varphi_k:\mathds{R}^n \to \mathds{R}$
real functions.
The joint probability function $f_{\vec{Y}}(\vec{y})$ is given by:
\begin{equation}
    \displaystyle
    f_{\vec{Y}}(\vec{y}) = 
    \int_{-\infty}^{\infty} d\vec{x} \,\,f_{\vec{X}}(\vec{x})
    \prod_{i=1}^m \delta\left( y_i - \varphi_i(\vec{x}) \right),
\end{equation}
%
where $\delta$ is the Dirac Delta function.
%
For the particular case of a unidimensional variable,
$X:\Omega\to\mathds{R}$ and
   $Y:\Omega\to\mathds{R}$, with $Y=\varphi(X)$,
\begin{equation}
    \displaystyle
   f_{{Y}}({y}) = 
   \int_{-\infty}^{\infty} d{x} \,\,f_{{X}}({x})
   \delta\left( y - \varphi({x}) \right)
\end{equation}   

Then, we can use this theorem to find the distribution of $\beta$
from $F_{X}(x)=U(0, 1)$ with the transformation:
\begin{equation}
\beta = \tan(acos(x))
\end{equation}
as well as from $f_{\Lambda}(\lambda)=sin(\lambda)$ with the transformation
\begin{equation}
\beta = \tan(\lambda), \quad 0<\lambda<\pi/2.
\end{equation}

Using the latter, we have:

\begin{align}
f_{B}(\beta) &= 
   \int_{-\infty}^{\infty} d{\lambda} \,\,f_{\Lambda}(\lambda)
   \delta\left( \beta - \tan(\lambda)) \right) \nonumber\\
   &=
   \int_{0}^{\pi/2} d{\lambda} \,\,sin(\lambda)
   \delta\left( \beta - \tan(\lambda)) \right) 
\end{align}

To solve this integral, we perform the change of variables:

$$
z=\tan(\lambda) \implies \lambda=atan(z), \quad d\lambda=\frac{dz}{1+z^2}
$$

Therefore,

\begin{equation}
   \displaystyle
   f_{B}(\beta) = 
   \int_{0}^{\infty} dz \; \frac{sin(atan(z))}{1+z^2} \;
   \delta\left( \beta - z \right) = \frac{\sin(atan(\beta))}{1+\beta^2}
\label{E_fB}
\end{equation}

This expression can be simplified using the properties of trigonometric
functions. Indeed, if $\beta= tan(z)$ for a number $z$, then:

\begin{align*}
\beta^{-2} + 1 & = \frac{1}{tan(z)^2} + 1 = \frac{\cos(z)^2}{\sin(z)^2} +1 = 
                \frac{\sin(z)^2+\cos(z)^2}{\sin(z)^2} \\
                & =\frac{1}{\sin(z)^2} 
\end{align*}
\begin{align*}
   \implies & \frac{1}{\sin(z)}= \sqrt{\beta^{-2} + 1} =
   \sqrt{\frac{1+\beta^{2}}{\beta^2}} = 
   \frac{\sqrt{1+\beta^2}}{\beta} \\
\implies & {\sin(z)}=\frac{\beta}{\sqrt{1+\beta^2}} \\
\implies & \sin(atan(\beta))=\frac{\beta}{\sqrt{1+\beta^2}} \\
\end{align*} 

Replacing in Eq.~\ref{E_fB} we have,

\begin{align}
   f_B(\beta)&=\frac{\sin(atan(\beta))}{1+\beta^2} \nonumber\\
   &=\frac{\beta}{\sqrt{1+\beta^2}}\,\,\frac{1}{1+\beta^2} \nonumber \\
   &= \beta (1+\beta^2)^{-3/2}
\end{align}

Therefore, the probability function is:

\begin{align}
F_B(\beta) &= \int_0^{\beta} f_B(b) d{b} \nonumber\\
           &= \int_0^{b} b (1+b^2)^{-3/2} d{b} \nonumber\\
           &= - \frac{1}{\sqrt{1+b^2}} \Bigg|_0^{\beta} \nonumber\\
           &= 1 - \frac{1}{\sqrt{1+\beta^2}}
\label{E_FB}
\end{align}
 
Figure~\ref{F_rvs_pdf}, top panel, shows the theoretical distribution of $\mathrm{\beta}$, with the Monte Carlo sampling of the random variable shown with the histogram.


Knowing the distribution $f_B$ one can perform analyses of the
orientations of vectors with respect to a particular direction.
Generally, it is not useful to measure a single value of the R.V. $\beta$, given that it is subject to random fluctuations.
Therefore, we calculate the values of the estimator $\beta$ in a sample
of observations. I. e., we analyze a random sample (R.S.) of values in order
to determine if it differs from the expected results for a random distribution
(a R.S. under the null hypothesis) of vectors.
To formalize these analyses we need to establish some basic properties 
of the distribution $f_B$.

%
%
%
%
%
%
%
%
%
%
%

%

The first moment of the distribution, if it exists, is:

\begin{align}
   E[B] &= \int_0^{\infty} t f_B(t) dt \nonumber \\
   &= \int_0^{1} t f_B(t) dt + \int_1^{\infty} t f_B(t) dt 
\end{align}

where
\begin{align*}
   &\int_1^{\infty} t f_B(t) dt 
   =
   \int_1^{\infty} t \frac{t}{(1+t^2)^{3/2}} dt 
   %
\end{align*}

Keeping in mind that for a real number $x>1$ we have $x^n>x$, and $x>\sqrt{x}$, therefore $x^{3/2}=x\sqrt{x}< x$. Then, for
$\beta>1$, $1+\beta^2>2>1$ and

   $$ \frac{1}{(1+t^2)^{3/2}} > \frac{1}{(1+t^2)} $$

Then we can limit the integral:
 
 \begin{align*}   
\int_1^{\infty} t \frac{t}{(1+t^2)^{3/2}} dt &> 
    \int_1^{\infty} \frac{t^2}{(1+t^2)} dt  \\
    &> \int_1^{\infty} \frac{t}{(1+t^2)} dt  \\
    &> \underset{M \to \infty}{lim}
   \left.\frac{1}{2}ln(1+t^2)\right|_1^{M} = \infty.
\end{align*}

\noindent
We see, then, that the expection value $E[B]$ is undefined.
%
Indeed, no moment of this distribution is defined. In fact, keeping in mind that:

\begin{align}
   E[B^n] &= \int_0^{\infty} t^n f_B(t) dt \nonumber \\
          &= \int_0^1 t^n f_B(t) dt  + \int_1^{\infty} t^n f_B(t) dt 
\end{align}
and that:
$$
\beta>1 \implies 1+\beta^1>1 \implies \frac{\beta^n}{1+\beta^2} >
\frac{\beta}{1+\beta^2}
$$
for $n\ge1$. Then,
\begin{align*}
   \int_1^{\infty} t^n f_B(t) dt &> \int_1^{\infty} t f_B(t) dt > \infty.
\end{align*}

The distribution $f_B(\beta)$ is a pathological distribution where the moments
are undefined.
The properties of this distribution are similar to the properties of the
Cauchy distribution.
This limitation prevents from using Monte Carlo procedures to estimate the
distribution of $\bar{\beta}$ because it is not possible to ensure that the
average values of $\beta$ follow a stable distribution.
In order to work with this distribution, we could devise a truncated distribution,
between arbitrary values $l1$ and $l2$, with the condition that $l1\sim0$ and $l2$ 
be much larger than the region of interest of the parameter $\beta$, which is the
region around $\beta=1$.

For example, if we choose
$$L_1=10^{-3}; \qquad L_2=10^3$$
%
it arises that, from defining the the correction factor $\kappa:$
\begin{align*}
\kappa =  \int_{L_1}^{L_2} f_B(t) dt = 
   \frac{1}{\sqrt{1+L_1^2}} -
   \frac{1}{\sqrt{1+L_2^2}} =
   \frac{1}{\sqrt{1+10^{-6}}} -
   \frac{1}{\sqrt{1+10^{6}}}
\end{align*}
%
we can define an approximation to the distribution function for $B$
defined as:
\[   
f_{\tilde{B}}(\beta) = 
\begin{cases}     
   \frac{1}{\kappa} f_B(\beta) & \qquad \beta \in [L_1, L_2] \\
0 &\qquad \mathrm{otherwise},
\end{cases}
\]

The mean of this function is in fact defined, and its expression is as follows:
\begin{align*}
   E[\tilde{B}] = \frac{1}{\kappa} \Bigg[ \Bigg.
   \frac{L_1}{\sqrt{L_1^2 + 1}}
 - \frac{L_1^2 asinh(L_1)}{L_1^2 + 1}
 + \frac{L_2^2 asinh(L_2)}{L_2^2 + 1} \\
 - \frac{L_2}{\sqrt{L_2^2 + 1}}
 + \frac{asinh(L_2)}{L_2^2 + 1} 
 - \frac{asinh(L_1)}{L_1^2 + 1} \Bigg. \Bigg]
\end{align*}

If we take $L_1=1/L_2$:

\begin{align*}
   E[\tilde{B}] = \frac{1}{\kappa} \Bigg[ \Bigg.
      \frac{1/L_2}{\sqrt{L_2^{-2} + 1}}
   - \frac{L_2^{-2} asinh(1/L_2)}{L_2^{-2} + 1}
   + \frac{L_2^{2} asinh(L_2)}{L_2^2 + 1} \\
 - \frac{L_2}{\sqrt{L_2^2 + 1}}
 + \frac{asinh(L_2)}{L_2^2 + 1} 
   - \frac{asinh(1/L_2)}{L_2^{-2} + 1} \Bigg. \Bigg]
\end{align*}

However, the result is strongly dependant on the value of $L_2$, and, to a lesser
extent, the value of $L_1$. Let

$$A(1/L_2, L_2) = \int_{1/L_2}^{L_2} t f_B(t) dt$$
%
It is straightforward to see that $A(1/L_2, L_2)$ 
depends on $L_2$, where we took different values for $L_2$ and $L_1 = 1/L_2$.

%
With this we prove that it is not possible to obtain a distribution for $\bar{\beta}$.
Furthermore, not only is it not possible to solve analytically, but it is also not 
possible to make a formal bootstrap estimation of the error.
However, as we show in the next section, $\beta$ can be used to define
a new parameter with better statistical properties.

\subsection{Test for the fraction of vectors in excess to random w.r.t. a reference direction}
\label{sec:eta}

To find a robust estimator we consider the fraction of values of $\beta$
that are greater than some critical value. Given that when the perpendicular
and parallel components are equal there is no preference in any of the two directions,
we can posit that said critical value be $\beta=1$. Therefore, we define the parameter:
\begin{equation}
    \hat{\eta} = \frac{N(\beta>1)}{N(\beta<1)}
\label{E_eta}
\end{equation}
%
where $N$ is the number of observations of a sample that fulfills the
conditions indicated between parentheses.
%
Under $H_0$, based on the probability density function, one expects that
\begin{equation}
    \eta_0 = \frac{P(\beta>1)}{P(\beta<1)}
\end{equation}

To calculate the value of $\eta_0$, we take into account that, 
using the probability function, $F_B$ (see Eq.~\ref{E_FB}):
$$P(\beta>1) = 1 - F_B(1) = 1 -
\left.\left(1-\frac{1}{\sqrt{1+\beta^2}}\right)\right|_{\beta=1}=\frac{1}{\sqrt{2}}$$
$$P(\beta<1) = F_B(1) =
\left.1-\frac{1}{\sqrt{1+\beta^2}}\right|_{\beta=1}=1-\frac{1}{\sqrt{2}}$$
i.e.:
\begin{align}
\eta_0 &= \frac{P(\beta>1)}{P(\beta<1)} = \frac{ \int_0^{1} f_B(t) dt }%
        { \int_1^{\infty} f_B(t) dt  } = \frac{\frac{1}{\sqrt{2}}}{1-\frac{1}{\sqrt{2}}} = \frac{1}{\sqrt{2}-1} \nonumber\\
       &\cong 2.4142
\end{align} 

\begin{figure}
   \centering
   \includegraphics[width=0.5\textwidth]{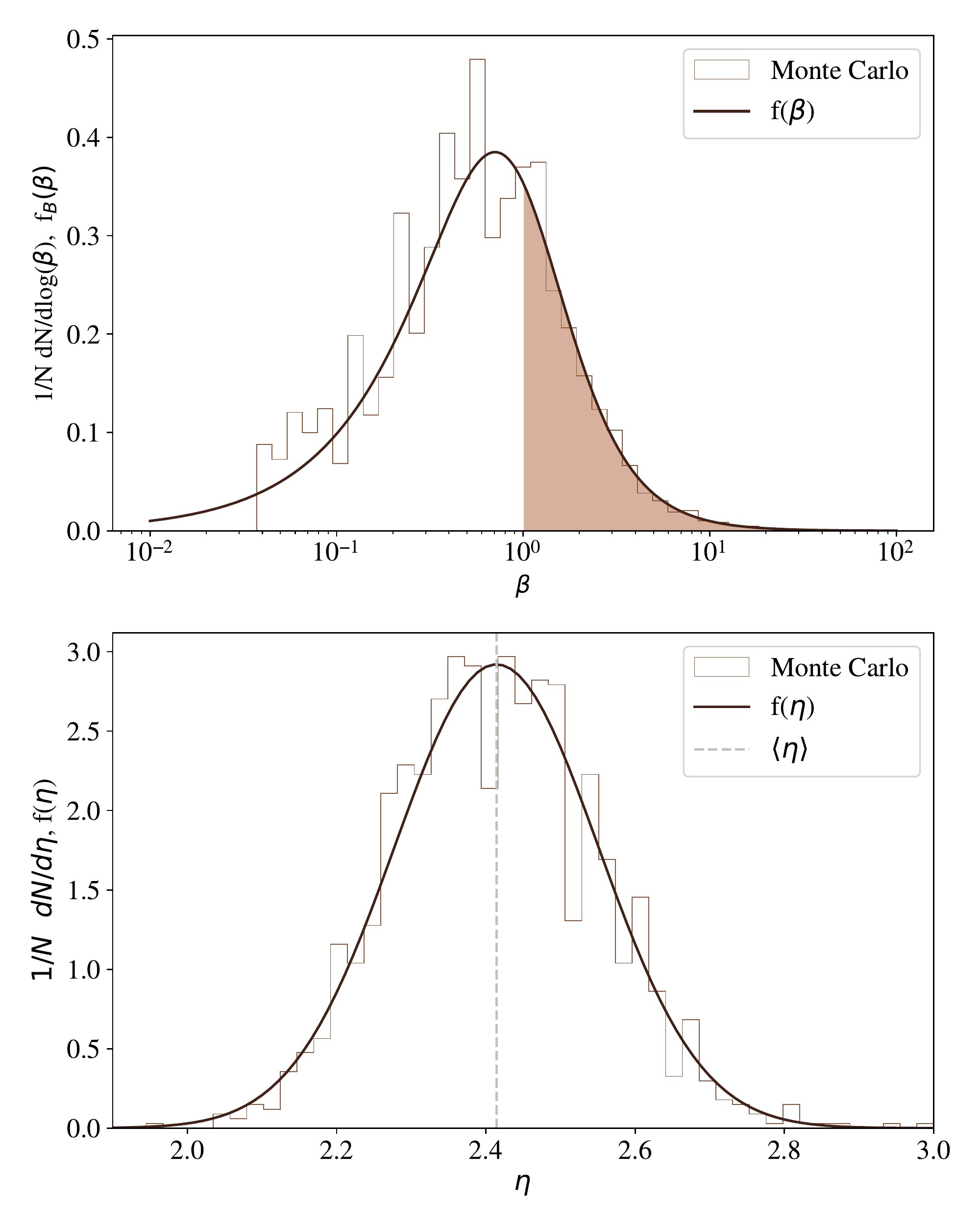}%
   \caption{(upper:) Distribution of $\beta$ obtained from the theoretical derivation (solid line) and from Monte Carlo simulations (histogram), and definition of $\eta$ (fraction of samples with $\beta>1$).
   (bottom:) Histogram of the $\eta$ variables sorted with
   the Monte Carlo method (from samples of $\beta$)
   and with the theoretical distribution approximation.
   The mean theoretical value ($1/\sqrt{2}$) is shown in the 
   dashed line and the histogram corresponds to a Monte Carlo
   realization of eta values, using 
   }
   \label{F_rvs_pdf}
\end{figure}



%
We propose that $\hat{\eta}$ is an estimator of $\eta_0$, i.e.,
we need to check if: $$ E(\hat{\eta}) = \eta $$
%
Let there be a random sample of $N$ values of $\beta$, we define:
$$n = N(\beta>1)$$
%
Given that the probability of obtaining a value of $\beta>1$ is
$P(\beta>1)=1/\sqrt(2)$, the variable $n$ has a binomial distribution,
$$
f_n(n) = Bin(p, N) = {N\choose n} p^n (1-p)^{N-n}
$$
with $p=1/\sqrt{2}\approx 0.707$.
Therefore, 
$$
\eta = \frac{n}{N-n}
$$
%
and the distribution of $\eta$ can then be calculated from the
distribution of $n$, taking into account that:

$$P_{\eta}\left(\eta=\frac{k}{N-k}\right) = P_n(n=k)$$

where $k=\frac{\eta N}{\eta - 1}$.

%
It is equivalent then, although much more efficient, to generate random 
variables of the distribution of $\eta$ with this method than
with a Monte Carlo method. The comparison between the two random samples can
be seen in the bottom panel of Figure~\ref{F_rvs_pdf}.
%

Then, taking into account the fact that the expectation value of the variable
$n \sim Bin(N, p)$ is $np$, we have to calculate the expectation value of the ratio.
This problem is generally not well defined, but it can be solved approximately.

Let X and Y be R.V. defined as $X=n$, $Y=N-n$. If $q=1-p$,
the expectation values of these variables are:
$$\mu_X = Np; \qquad \mu_Y=N-Np = N - \mu_X$$
and the variances:
$$\sigma_X^2 = \sigma_Y^2 = NP(1-p) = Npq$$
with $q=1-p$.

\begin{figure}
\includegraphics[width=0.5\textwidth]{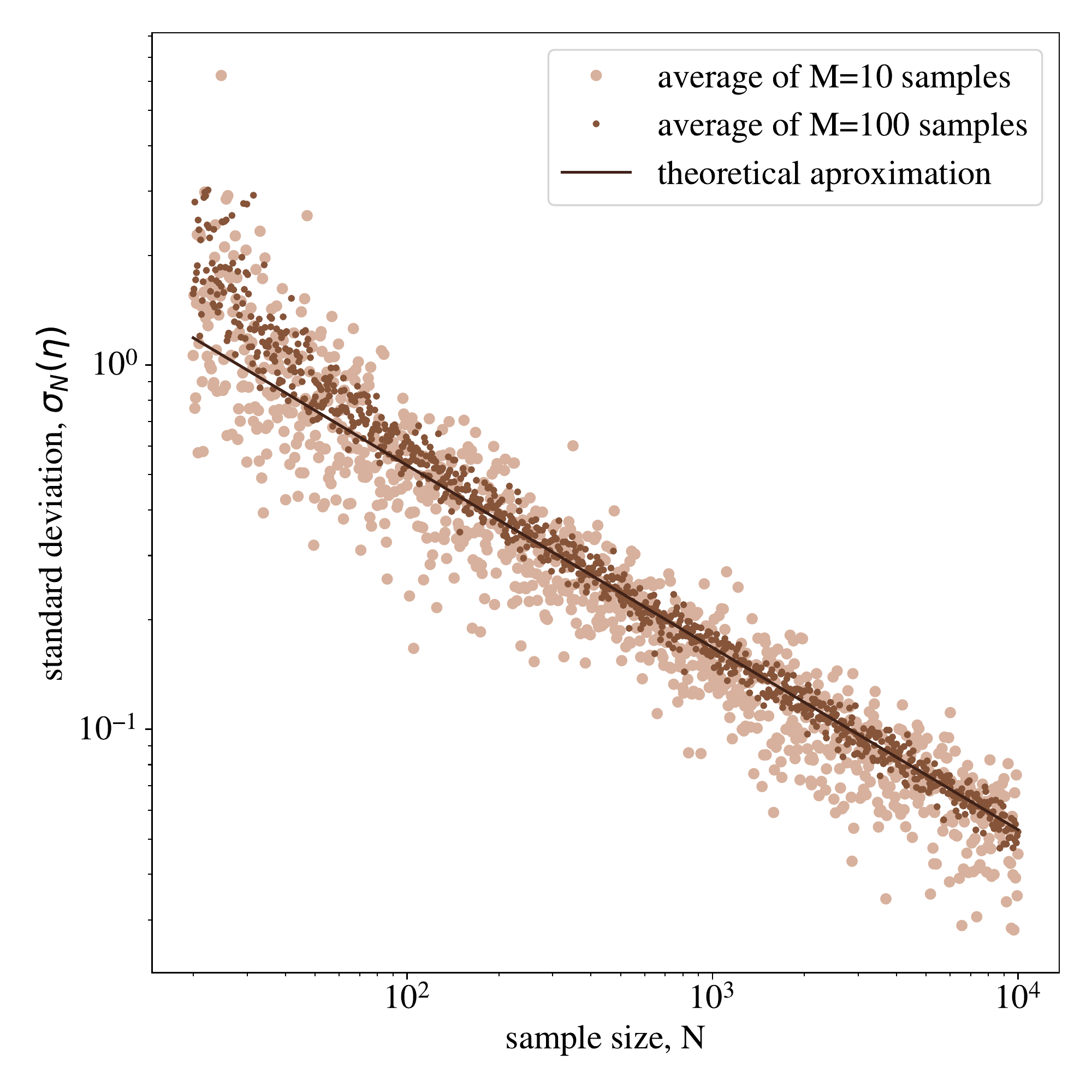}
\centering
   \caption{Variation of Monte Carlo estimations of the variance of M samples of values
   of $\eta$, calculated from N samples of values of $\beta$ (N, in the X axis, simulated), for M=10 (large dots) and M=100 (small dots). The theoretical variance is shown to be an appropriate estimation for samples with size M>100.}
\label{eta_variance}
\end{figure}

In the Appendix we show that the variance of this estimator is given by

\begin{equation}
    Var(\eta)\simeq\frac{28}{N},
\end{equation}

\noindent
and Fig.~\ref{eta_variance} shows that for sample sizes larger than approximately 100, using the theoretical value is equivalent to using Monte Carlo simulations, with the advantage of needing comparatively no computation time.

\subsection{A test for the OLS coefficients of the cosine distribution}
\label{S_lambda}

Another option is to analyze the distribution of cos$(\lambda)$ to
determine if it is distinguishable from the expected distribution
of a sample of random orientations of vectors $\vec{S}$.
As discussed, said distribution is uniform under $H_0$.
Working with samples of limited size, the statistical fluctuations
can generate differences between the two sets of data, even when they
arise from the same distribution.
Therefore, we want to compare the two distributions and establish whether
their difference is sufficient to discard $H_0$.
            
The comparison between two observed distributions is performed in a more
robust way from the empirical cumulative distribution. If one has a random sample
of a variable $X$, $\{x_i\}_{i=1}^n$, where the values are sorted,
the empirical cumulative distribution function (ECDF) is:

\begin{equation}
F_e[x] = \frac{|\{X / X< x\}|}{|\{X\}|}
\label{E_ecdf}
\end{equation}


This function is used in the Kolmogorov-Smirnov test, where
the statistic $D$ is defined as the maximum difference between the two
cumulative distributions.
Following this idea, to describe the difference between an observed 
distribution and a control distribution $F_c(x)$, we consider:

\begin{equation}
\Delta(x) = F_e[X](x) - F_c(x)
\end{equation}

In the case of the distribution of the $\lambda$ parameter, we know that
under $H_0$ it has to be uniform between $0$ and $\pi/2$, which yields 
$F_c(x) = 2x/\pi$. Then,

\begin{equation}
\Delta(x) = F_e[X](x) - \frac{2x}{\pi}
\label{E_resEcdf}    
\end{equation}

This function, by definition, begins and ends in zero, i.e.,
$$
\Delta(x)=0 \textrm{ for }x=0;\, x=\pi/2.
$$
%

                   

\begin{figure}
  \centering
\includegraphics[width=\columnwidth]{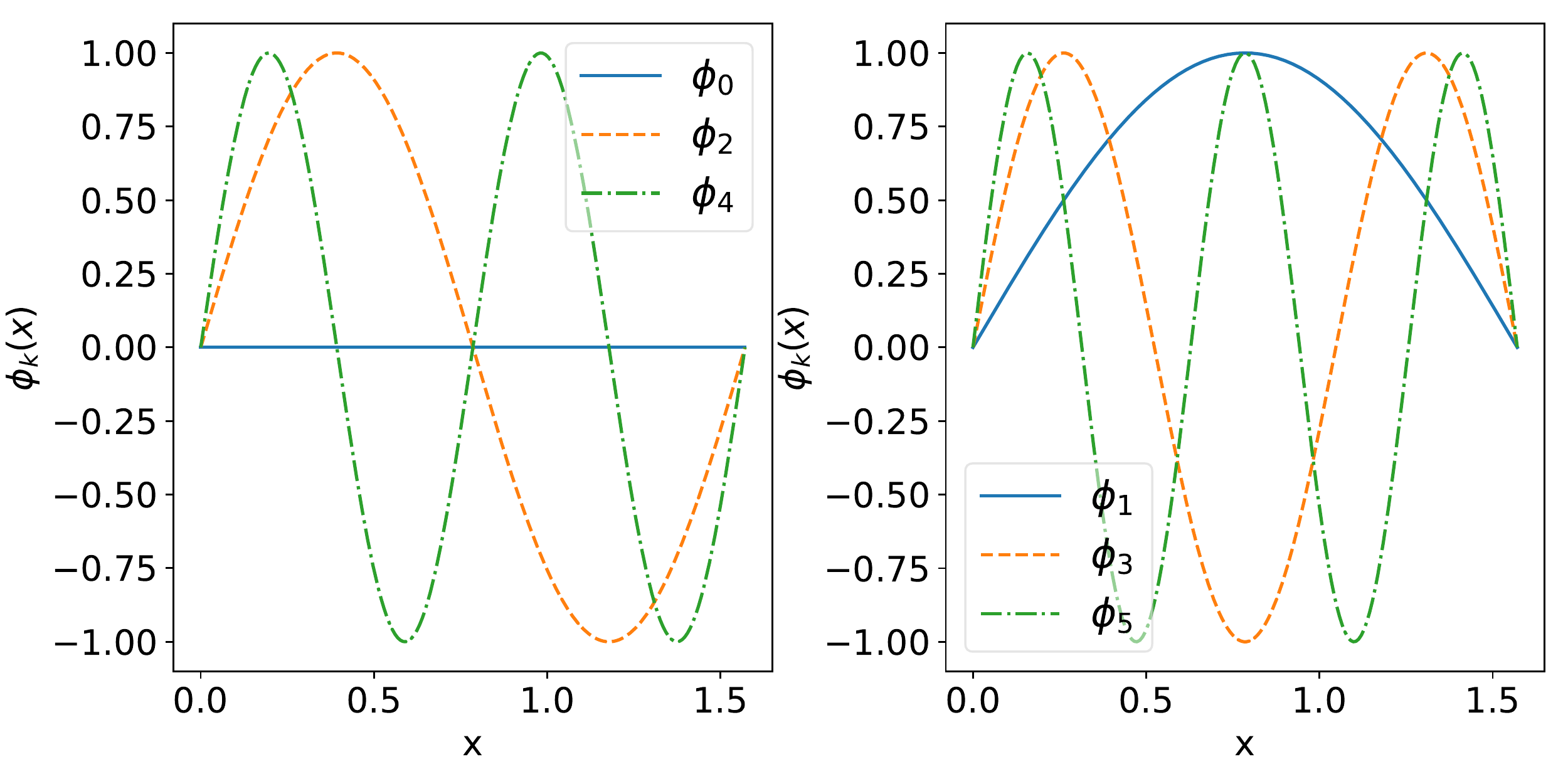}
\caption{First even (left panel) and odd (right panel) elements of the base of functions $\phi(x)$.}
  \label{F_sinbasis}
\end{figure}

In order to represent the function $\Delta(x)$ one can use a base of
orthogonal functions. If $f(x)$ is a continuous function in the interval 
$[0, \pi/2]$, then it can be written as:

\begin{equation}
\Delta(x) = \sum_{k=0}^{\infty} a_k \phi_k(x)
\end{equation}
where
$$
\phi_k(x) = sin(2kx).
$$
%
The first elements of this base can be seen in Figure~\ref{F_sinbasis}.
%
%
%
%
%
%
%
%
%
%
%
%
%
%
And by defining orthogonality of the base we can finally write our orthonormal system as:

\begin{equation}
    \phi_k(x) = \frac{4}{\pi}sin(2kx).
\end{equation}

That said, in the case of data sets, there is no continuous function, but a 
discrete sampling.
Fourier analysis allows for an expansion in terms of a finite sum of sines
in the case of a set of equally distanced points.
If the points are not equally distanced we must resort to alternative strategies.
One might make a binning, but this alternates the information available.
Another option is to fit a model to the observed points.
For example, we might take as a model:

\begin{equation}
\Phi(x) = \sum_{k=0}^M a_k \phi_k(x),
\label{E_armonic}
\end{equation}

\noindent which is a model that is linear in the parameters that we want to calculate:
the coefficients $a_k$.



%
Even though this method is enough to represent a distribution, it is not
useful as a statistic to characterize the parameters of the distribution.
However, it is possible to take advantage of the conditions of symmetry
to characterize the distribution.

For example, if we assume symmetry, i.e. consider only the cosines of acute angles, we expect the even parameters to be zero.
%
%
Ec.~\ref{E_armonic} is modelling the difference between a cumulative distribution of the cosine distribution calculated from data and a control function, so it is expected that when studying the effects of preferential alignments in a vector population this function will take values that are either mostly positive or negative, indicating a net alignment perpendicular or parallel to the preferred direction, respectively.
Therefore, in the case of symmetry where one is interested in the acute angle $\mathrm{\lambda}$, it is noticeable from the right panel of Fig.~\ref{F_sinbasis} that the first term, represented in a blue curve, will be the dominant term in Ec.~\ref{E_armonic} and is proportional to the parameter \textit{$a_\mathrm{1}$}; the area under the curve, whether positive or negative, will
represent a net alignment in the perpendicular or parallel direction, respectively. 

The coefficients $a_\mathrm{k}$ can be obtained from ordinary least squares from data. Given Ec.~\ref{E_armonic} and the residuals of the ECDF of the data $\mathrm{y_i=i/n-i}$ of size $n$, we intend to minimize:

\begin{equation}
\chi^2 = \sum_{i=1}^{n} \left(\frac{y_i}{\sigma_i} - \frac{1}{\sigma_i} \Phi(x) \right)^2 = |\mathbf{A} a_k - \mathbf{B}|,
\end{equation}

\noindent where \textbf{A} and \textbf{B} are in matrix notation and $x$ is the data we want to fit, i.e. the cosines calculated from the sample. So by minimizing this expression it follows that $\mathbf{A}a_\mathrm{k}$ equals \textbf{B} \citep[Chapter 3]{hastie01statisticallearning}{}, so:

\begin{equation}
a_k = \frac{(\mathbf{A}^T \mathbf{B})_k}{(\mathbf{A}^T \mathbf{A})_k}.
\end{equation}

The base of functions is the set of all harmonic functions $\mathrm{\phi_k(x)=sin(k\pi x)}$, which we truncate at k=4, and assuming $\mathrm{\sigma_i=1}$, we have:

\begin{equation}
(\mathbf{A}^T \mathbf{A})_k = \sum_{i=1}^{n} \phi_k^2(x_i).
\end{equation}

On the other hand, we have $(\mathbf{A}^T \mathbf{B})_k = \sum_{i=1}^{n} \phi_k(x_i) y_i$, so, after replacing we have:

\begin{equation}
a_k = \frac{\sum_{i=1}^{n} \phi_k(x_i) y_i}{\sum_{i=1}^{n} \phi_k^2(x_i)}.
\end{equation}

In our case of study we have $\mathrm{x=cos(\lambda)}$, and assuming sorted data:
$\mathrm{y_i=i/n-i}$, we finally arrive at an analytical expression for the OLS coefficients:

\begin{equation}
    a_k = \frac{\sum_{i=1}^{n} i\,sin[k\pi cos(\lambda_i)]}{\sum_{i=1}^{n} sin^2[k\pi cos(\lambda_i)]} \left( \frac{n-1}{n} \right).
\end{equation}
This is an expression that directly relates the parameters to the data. As previously indicated, the coefficient $a_\mathrm{1}$ (Ec.~\ref{E_a1}) gives the first order approximation for the residual function (Ec.~\ref{E_armonic}): 

\begin{equation}
    a_1 = \frac{\sum_{i=1}^{n} i\,sin[\pi cos(\lambda_i)]}{\sum_{i=1}^{n} sin^2[\pi cos(\lambda_i)]} \left( \frac{n-1}{n} \right).
\label{E_a1}
\end{equation}

If the data consists of a vector population with a net alignment perpendicular to the preferred direction, Ec.~\ref{E_armonic} will resemble the blue curve of the right panel of Fig.~\ref{F_sinbasis} with the coefficient $a_\mathrm{1}$ taking positive values. This is due to the data presenting an excess in the lower values of cosines and, as a consequence, the ECDF taking values larger than the control function so that the residues are positive. On the other hand, if the data presents a net alignment in the parallel direction, $a_\mathrm{1}$ will take negative values.

\section{Application to synthetic data and comparison with other methods}
\label{S_Comparison}

To test the efficiency of the methods presented above with regards to usual methods, such as the average cosine, we apply them to 3 sets of synthetic data. These data are generated sorting random points on the surface of a 3--dimensional ellipsoid with axis a, b, and c, with various eccentricities defined in the usual manner: $e^2 = 1-c^2/a^2$, where $c<a$ and $a=b$. It is worth noting that by varying the c axis, we are defining this vertical $z$ direction as the preferential direction for spherical symmetry. We chose to establish three different eccentricities to test the methods: $e^2$ = 0.6, 0.4, and 0, going from elongated to isotropic, respectively. In this manner, we are simulating a population of vectors with no preferential orientation for the isotropic case, to one with a strong alignment trend for the largest eccentricity.

First, we studied the stability of the estimators with various sample sizes. Figure~\ref{F_methods_v_N} shows the mean and standard deviation calculated for 50 random realizations of samples of size $\mathrm{N_{ran}}$. For a sample size of over a few hundred the estimation of the parameters $a_\mathrm{1}$ and $\eta$ is reliable, and for sample sizes of over $\mathrm{\sim10^4}$ the relative error is sufficiently small to distinguish between little variations in eccentricities. This is promising in the sense that, for present and future large scale surveys with large samples, even a small effect of alignment would be detectable with these methods.

\newcommand{\Nbs}{50 }

\begin{figure*}
   \centering
   \includegraphics[width=.8\textwidth]{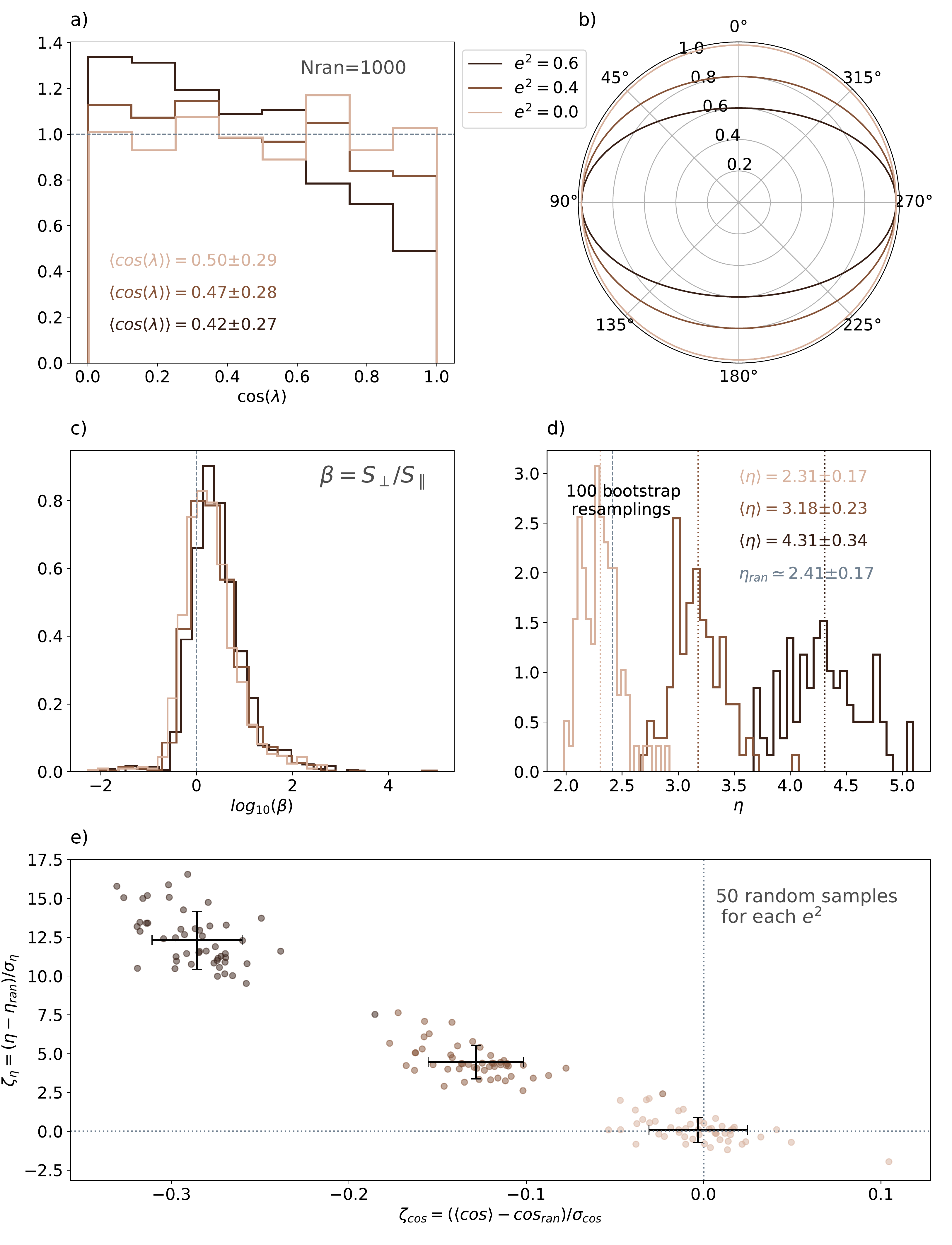}
   \caption{Representation of the method starting from a distribution of alignments of a vector population arriving to the ``fraction of vectors'' parameter $\eta$, along with its significance compared to the average of cosines. 
   Panels a) and b) show the histograms of cosines of angles, which is the usual manner of studying alignments, corresponding to directions scattered along the surface of ellipsoids with eccentricities 0.6, 0.4, and 0, i.e. from strong alignment to random behaviour. The first panel includes the mean and standard deviation of the distribution of cosines.
   The method then consists of calculating the parameter $\beta$ as the ratio of perpendicular to parallel components, whose distributions for the three alignments are shown on panel c), and then obtaining $\eta$: the number of vectors with $\beta>1$. Sample variance is taken into account by bootstrap resampling the data, and thus a mean and standard deviation for $\eta$ can be estimated (panel d). To assess the stability of the parameter we perform this calculation several times by varying the random seed of the initial samples and thus obtaining several estimations of the bootstrap mean of $\eta$. 
   And finally, in order to quantify the significance of the estimated alignment signal, we define a variable $\zeta_\mathrm{\eta}$ normalized with respect to the behaviour of the estimator $\eta$ under the null hypothesis. Panel e) shows $\zeta_\mathrm{{\eta}}$ against $\zeta_\mathrm{{\langle cos\rangle}}$, where we find that $\eta$ can detect alignment signal with higher statistical significance. }
   \label{F_zetaeta_vs_zetacos}
\end{figure*}

In Fig.~\ref{F_zetaeta_vs_zetacos} we test the statistical significance of the $\eta$ parameter when compared to the average of cosines, $\mathrm{\langle cos(\lambda)\rangle}$. As explained above, we generate random points along the surface of three ellipsoids with eccentricity values of 0.6, 0.4, and 0. These populations yield the three cosine histograms showed in panel a), where we include the mean values along with the standard deviation of the distribution. We note that the standard deviation of the cosine distribution is of the order of its mean.

Panel c) shows the logarithm of the $\beta$ parameter defined as the ratio of perpendicular and parallel components of the vector, the parallel direction being that of the preferential direction for spherical symmetry. The mean value of $\beta$ is similar for every population, so this is not an ideal parameter to study. However, the cumulative number of vectors that have a larger perpendicular component, a.k.a. the $\eta$ parameter, is noticeable. 

In order to account for sample variance, we perform bootstrap sampling of our observable $\beta$ to obtain a distribution of $\eta$ from which we can define a mean and a standard deviation. Such bootstrap distributions for each eccentricity are shown in panel d). We include the mean value and standard deviation of the distributions, as well as that of the isotropic case in grey colour, which has been derived theoretically in Sec.~\ref{sec:eta}.

It is noticeable that the $\eta$ distributions have larger standard deviation for larger mean values. This is a consequence of the definition of $\beta$, where values for larger parallel components are limited between 0 and 1, while values for larger perpendicular values have no theoretical upper limit. If one were to define $\beta$ in the inverse manner, the same divergent behaviour happens for larger parallel alignments.
This is a feature of the parameter to keep in mind. However, while the upper limit is infinite in theory, it is not in practice. On one hand one would have to find vectors of infinite norm for this to be a problem. Furthermore, alignment corresponding to an eccentricity of 0.6 is unlikely to be found in observables such as galactic orientations, much less higher values of eccentricity. In other words, we are testing these parameters in the limits of practical situations.

To finally assess the efficacy and significance of the $\eta$ parameter, we repeat the above procedure by generating random points along the surface of the different ellipsoids with \Nbs  random seeds, therefore yielding \Nbs values of $\eta$ and average cosines. Furthermore, to study the statistical significance with respect to random behaviours we define the variable $\zeta$ as:

\begin{equation}
    \zeta_X = \frac{X-\Bar{X}_{ran}}{\sigma_{X,ran}},
\end{equation}

\noindent
where X is the random variable we want to test, in this case: $\eta$ and $\mathrm{\langle cos(\lambda)\rangle}$. Given its definition, the $\zeta_\mathrm{X}$ variable contains information, not only about how much does the variable X deviates from isotropic behaviour, but also how statistically significant this deviance is. The isotropic values for the mean and standard deviation for $\eta$ have been theoretically derived in Sec.~\ref{sec:eta}.
The mean and standard deviation for the cosines in the isotropic case can be calculated as those of a uniform distribution. The probability distribution function of a uniform distribution is:

\begin{equation}
  f(x) =
    \begin{cases}
      \frac{1}{b-a} & \text{for a $\leq x\leq$ b}\\
      0 & \text{otherwise}
    \end{cases}       
\end{equation}

\noindent
where, in the case of the cosines, a=0 and b=1. So the mean and standard deviation would be:

\begin{equation}
    E(\mathrm{cos_{ran}}) = 0.5,
\end{equation}
and
\begin{equation}
    \sigma_{\text{cos, ran}} = \sqrt{\frac{1}{12}} \simeq 0.289. 
\end{equation}

Panel e) shows the computation of $\zeta_\mathrm{\eta}$ and $\zeta_\mathrm{{cos}}$. We observe that the statistical significance for deviations from isotropic behaviour as measured with $\eta$ is much higher than with the cosines. For an eccentricity of $e^2=0.6$ we have a significance of around 12.5 for $\eta$ and 0.3 for the cosines.  

\begin{figure*}
   \centering
   \includegraphics[width=\textwidth]{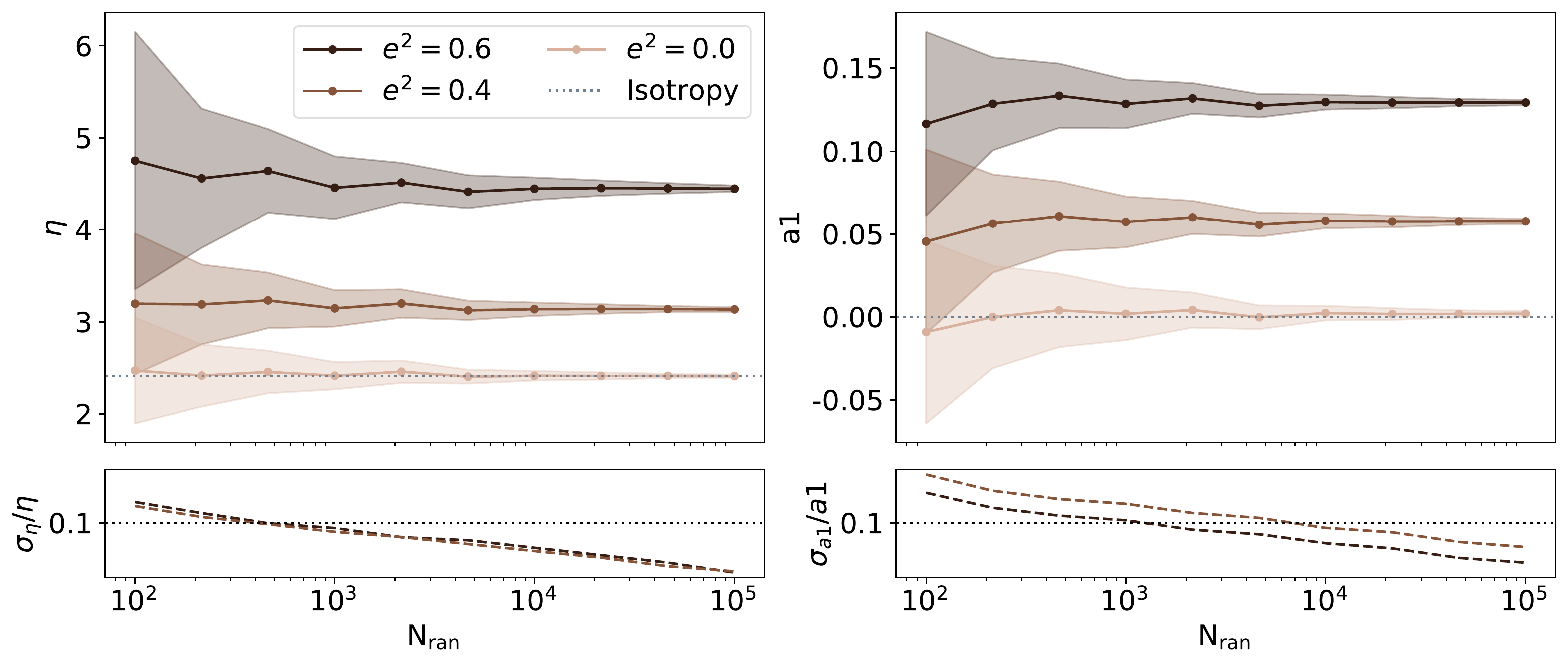}
   \caption{Stability of the methods with respect to the sample size. We applied the methods to \Nbs random realizations of synthetic data with alignments corresponding to three values of increasing eccentricities: 0, 0.4, and 0.6. The mean and standard deviations, represented by the solid lines and shadowed regions respectively, were calculated with these \Nbs independent results. We find that the mean of the parameters $\eta$ and $a_\mathrm{1}$ is stable even with a sample size of a few hundred. A relative error of 10\% is achieved with a sample size of $\sim10^3$ for the $\eta$ parameter. The same sample size for the same relative error is achieved when applying the OLS coefficient method to the data with the larger eccentricity.}
   \label{F_methods_v_N}
\end{figure*}

\begin{figure*}
   \centering
   \includegraphics[width=.8\textwidth]{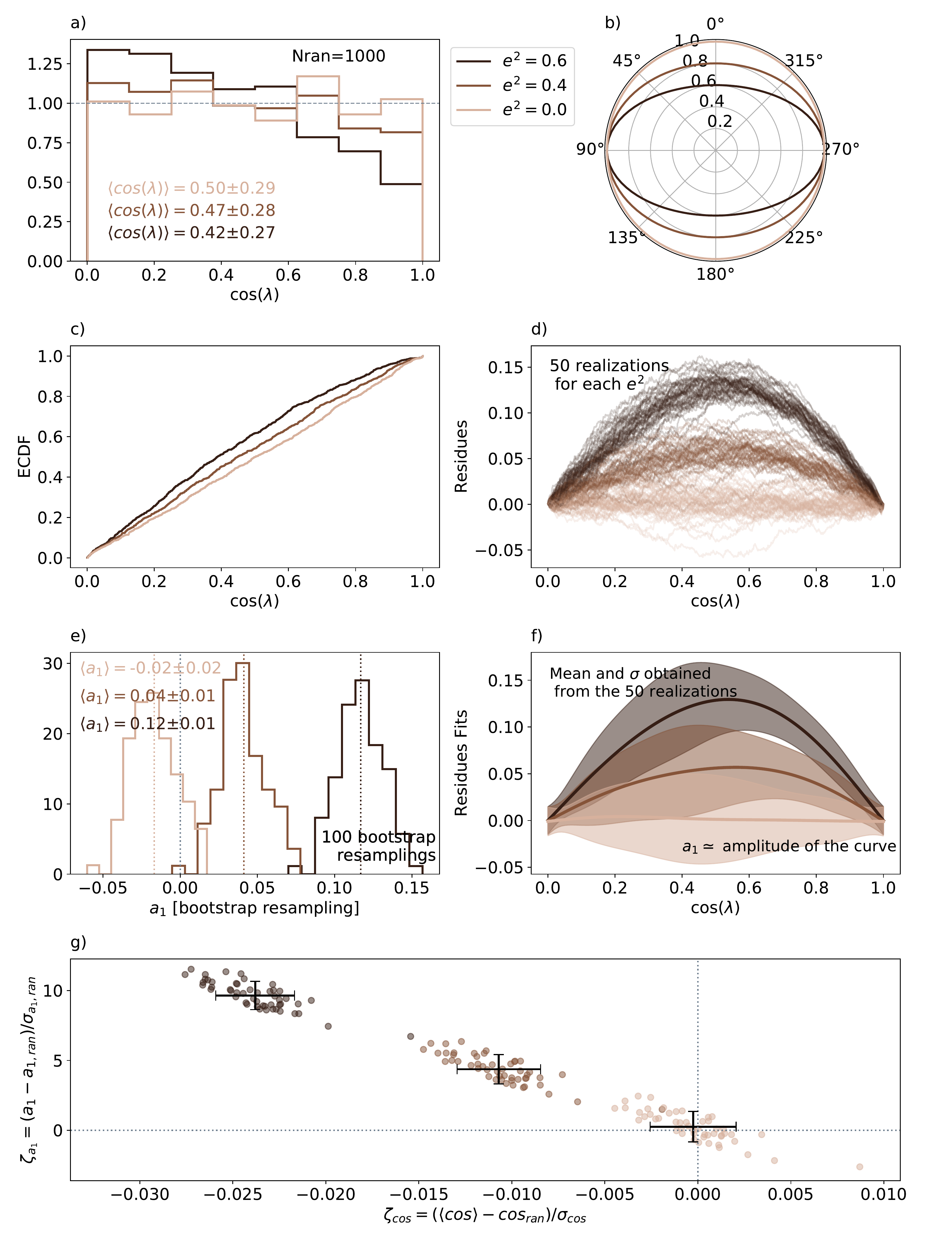}
   \caption{Representation of the method starting from a distribution of alignments of a vector population and deriving the the OLS coefficient $a_\mathrm{1}$, along with its significance compared to the average of cosines.
  Panels a) and b) show the histograms of cosines of angles, which is the usual manner of studying alignments, corresponding to directions scattered along the surface of ellipsoids with eccentricities 0.6, 0.4, and 0, i.e. from strong alignment to random behaviour. The first panel includes the mean and standard deviation of the distribution of cosines.
  The method then consists of calculating the residues of the ECDF of the cosine distribution, shown in panel c), and that of an isotropic behaviour. To assess the stability and significance of the parameter we perform this calculation several times by varying the random seed of the initial samples and thus obtaining the residual curves shown in panel d).
  Fits are performed over these curves. The amplitudes of said fits are characterized by the parameter $a_\mathrm{1}$, whose distributions along with their mean and standard deviation are shown in panel e). A representation of the fits is plotted in panel f), with a solid line representing the mean and a colored band showing their standard deviation. 
   %
   And finally, in order to quantify the significance of the estimated alignment signal, we define a variable $\zeta_\mathrm{a_1}$ normalized with respect to the behaviour of the estimator $a_\mathrm{1}$ under the null hypothesis. Panel g) shows $\zeta_\mathrm{{a_1}}$ against $\zeta_\mathrm{{\langle cos\rangle}}$, where we find that the coefficient $a_\mathrm{1}$ can detect alignment signal with higher statistical significance. 
  }
   \label{F_zetaa2_vs_zetacos}
\end{figure*}

In Fig.~\ref{F_zetaa2_vs_zetacos} we perform an equivalent analysis for the OLS coefficients method using the same synthetic data, as can seen by comparing the cosine distributions in panels a) of both Fig.~\ref{F_zetaa2_vs_zetacos} and \ref{F_zetaeta_vs_zetacos}. 
For this method we first calculate the ECDF of the cosines (panel c) of the data corresponding to the three cases of varying anisotropy. The residues are calculated by substracting from the ECDF of the data the one corresponding to an isotropic distribution which is the straight line of ECDF(cos($\lambda$)) = cos($\lambda$). 
We generate the data and perform this calculation \Nbs times with different random seeds, as shown in panel d). We fit each of this curves and plot the mean of the fit with solid lines and their corresponding 3$\sigma$ with the shadowed bands in panel f). 

The linear regression for each curve yields a set of coefficients $a_\mathrm{k}$, where the one that determines the basic shape of the fit is $a_\mathrm{1}$ (see section~\ref{S_lambda}). We perform bootstrap resampling of the data in order to estimate the mean and standard deviation of this coefficient. Panel e) shows the bootstrap distribution of the $a_\mathrm{1}$ coefficient corresponding to the cosine distributions shown in the same colors, where the dotted vertical lines correspond to the mean. Furthermore, in this panel we indicate in text the mean values along with the standard deviation of the distribution. It is readily noticeable that the mean of this coefficient is more robustly determined than the mean of the cosines.

Finally, for each of the random realizations we plot the normalized parameters $\zeta_\mathrm{a_1}$ and $\zeta_\mathrm{{\langle cos \rangle}}$, were we find that the OLS coefficient detects alignment with higher statistical significance.

\section{Summary}
\label{S_Summary}

In this paper we present two methods to detect and quantify alignment signals and test their statistical robustness. The first method uses the deviance of the relative fraction of vector components in the plane perpendicular to a reference direction with respect to an isotropic distribution. 
We have derived the first and second moments of the distribution of the resulting estimator, $\eta$, and can thus reliably assess its statistical significance. 
The second method is based on a fit over the residuals of the ECDF of the data with respect to the one expected for a uniform distribution. The fit uses a small set of harmonic orthogonal functions and does not rely on any binning scheme. The amplitude of the fit, i.e. the amplitude of the alignment signal, can be described by the first odd OLS parameter, $a_\mathrm{1}$. 

For the first method, we derive the distribution of the test statistic $\beta$, defined as the ratio between the perpendicular and parallel vector components (Eq.~\ref{E_beta}), under the null hypothesis.
We find that the probability distribution $f_B(\beta)$ is a pathological distribution where the moments are undefined, and as such, is not a robust statistic.
However, using this statistic we consider the fraction of its values that are greater than one, given that when the perpendicular and parallel components are equal there is no preference in any of the two directions and so $\beta=1$ can be taken as a critical value. Therefore, we define this ratio as the parameter $\eta$ in Eq.~\ref{E_eta}.
We find that, for $\beta$ defined as in Eq.~\ref{E_beta}, the parameter $\eta$ has an expectation value and variance given by $\eta_0\simeq$~2.4142 and Var$(\eta)\simeq28.1421/N$, respectively. The gaussian behaviour of this parameter allows for the first two moments to be sufficient to describe its distribution. The advantage of knowing the theoretical distribution of the parameters is the ability to accurately determine the statistical significance of any signal detected without investing computation resources and time into Monte Carlo simulations.

The second method of alignment analysis we presented yields OLS coefficients of a fit of the residues of the ECDF of the cosines of the data with respect to a random sample. The first odd coefficient, $a_\mathrm{1}$, of the harmonic expansion of the residual function is sufficient to characterize the amplitude of the alignment signal, with zero being consistent with isotropic orientations. Positive values of $a_\mathrm{1}$ indicate perpendicular alignment while negative values indicate parallel alignment with respect to the preferred direction.

We have compared these methods with others commonly used in the literature, mainly based on the average of the cosine distribution and using Monte Carlo simulations. This comparison was achieved by 
simulating a population of vectors with three different degrees of alignment (from no alignment, to intermediate, to greatly aligned), and testing how well the alignment signal was recovered by both the new and traditional parameters.

We find that the proposed statistics allow the detection of alignment signals with a larger significance. For a deviation of approximately  0.25$\sigma$ from an isotropic distribution of cosines, we obtain a significance of 10-- and 12$\sigma$ for the OLS coefficient $a_\mathrm{1}$ and the $\eta$ parameter respectively. In a forthcoming paper (Dávila-Kurbán et al., in prep) we apply the first method presented in this paper, i.e. the fraction of vectors parameter $\eta$, to data in a cosmological simulation. 


We have assessed the effects of uncertainties in the measurement of parallel and perpendicular vector components. 
To that end, we modelled observational errors by introducing gaussian noise into the components. The standard deviation was chosen to be 10 per cent of the mean vector norm. 
There is a linear relation between the "real" parameters, $\eta$ and $a_\mathrm{1}$, obtained with the raw synthetic data, and the "observed" parameters that account for the introduced mock observational error.
We performed 100 realizations of the calculations with and without mock errors.
The y--intercept \textit{b} corresponding to the maximum divergence from the real values can be expressed, for a sample size \textit{N} of 1000 and 5000 respectively, as $\mathrm{\Delta b_{\eta} = .7}$ and .3 for the first method, and $\Delta b_\mathrm{a_1}$ = .03 and .01 for the second method.
For reference, the range of values of the two parameters for the three eccentricities tested are: $\mathrm{2.0 < \eta < 5.5}$ and -.05 $<a_\mathrm{1}<$ 0.16 for N=1000; and 2.2 $<\eta<$ 5.0 and -.02 $<a_\mathrm{1}<$ 0.15 for N=5000.
This can be used to roughly estimate the size of samples required to achieve the detection of small signal in the data. With these tools, large forthcoming surveys and simulations can provide new insights on small amplitude signals of alignments unseen in current surveys with lower number of galaxies.

\section*{Acknowledgements}

This work was partially supported by the Consejo Nacional
de Investigaciones Cient\'ificas y T\'ecnicas (CONICET, Argentina), the Secretar\'ia de Ciencia y Tecnolog\'ia, Universidad Nacional de C\'ordoba, Argentina, and the Agencia Nacional de Promoci\'on de la Investigaci\'on, el Desarrollo Tecnol\'ogico y la Innovaci\'on, Ministerio de Ciencia, Tecnolog\'ia e Innovaci\'on, Argentina.
%
%
This research has made use of NASA’s Astrophysics Data System. Visualizations made use of python packages and inkscape software.

\section*{Data Availability}

The data underlying in this article are available on request to the corresponding author.

\clearpage
\bibliographystyle{mnras}
\small
\bibliography{biblio}

\onecolumn
\appendix

\section{Variance of $\hat{\eta}$}

We want to calculate the expectation value for the ratio $Q=X/Y$.
From the definition of the expectation value it stems that it is 
not possible to derive a simple expression for $Q[R]$.
Another reason that prevents from analytically solving the distribution 
of $Q$ is that the denominator can be zero. A way of solving this problem
is to rewrite the function in a way that avoids having a singularity.
It is possible to make such an approximation from developing the Taylor series
of $\bar{Q}(X, Y)=X/Y$ around $(X,Y)=(\mu_X, \mu_Y)$, i.e. $\bar{Q} = Q + R$:

\begin{eqnarray*}
\bar{Q}(X, Y) &= \bar{Q}(\mu_X, \mu_Y) 
          + \frac{\partial}{\partial X} \bar{Q}(\mu_X, \mu_Y) (X-\mu_X) \\
          &+ \frac{\partial}{\partial Y} \bar{Q}(\mu_X, \mu_Y) (Y-\mu_Y)
          + R,
\end{eqnarray*}
%
where $R$ is the order 2 error given by the Taylor theorem
\citep{Duris2018, koopman_confidence_1984, price_conficence_2008}.
%
Then, we can estimate the expectation value of $\bar{Q}$, which is
approximately $E[\bar{Q}] \approx E[Q]$, where: 

\begin{align*}
   E[Q] =& E\Big[Q(\mu_X, \mu_Y)
        +  \frac{\partial Q}{\partial X} (\mu_X, \mu_Y) (X-\mu_X)
        + \frac{\partial Q}{\partial Y} (\mu_X, \mu_Y) (Y-\mu_Y) 
        \Big] \\
        =& E\Big[Q(\mu_X, \mu_Y)\Big] 
        +  E\Big[\frac{\partial}{\partial X} Q(\mu_X, \mu_Y) (X-\mu_X)\Big]
        + E\Big[\frac{\partial}{\partial Y} Q(\mu_X, \mu_Y)
        (Y-\mu_Y)\Big] \\
        =& Q(\mu_X, \mu_Y)
        + \frac{\partial Q}{\partial X}(\mu_X, \mu_Y) E\Big[(X-\mu_X)\Big]
        + \frac{\partial Q}{\partial Y}(\mu_X, \mu_Y)
        E\Big[(Y-\mu_Y)\Big] \\
        =& Q(\mu_X, \mu_Y)
\end{align*}

Then, to a first order approximation and considering $Q=\eta$, $X=n$, $Y=N-n$:

\begin{align*}
   E(\hat{\eta}) &\approx \frac{Np}{N-Np} \\
\end{align*} 
and 
$$\frac{Np}{N-Np} = \frac{p}{1-p} = 
\frac{\frac{1}{\sqrt{2}}}{1-\frac{1}{\sqrt{2}}} =
\frac{P(\beta>1)}{P(\beta<1)} = \eta_0$$
%
therefore
$$
E(\hat{\eta})\approx \eta_0
$$

The second order moment of $\hat{\eta}$ can be obtained calculating
the variance of the first order approximation of $Q(X,Y)$:

\begin{align*}
Var(Q) &= Var\Big[
           Q(\mu_X, \mu_Y)
           +  \frac{\partial Q}{\partial X}(\mu_X, \mu_Y) (X-\mu_X)
           + \frac{\partial Q}{\partial Y}(\mu_X, \mu_Y) (Y-\mu_Y)
           \Big]\\
       &= 
         \left(\frac{\partial Q}{\partial X}(\mu_X, \mu_Y)\right)^2 Var[X-\mu_X] 
        + \left(\frac{\partial Q}{\partial Y}(\mu_X, \mu_Y)\right)^2 Var[Y-\mu_Y] 
        + \\
        &\qquad+\left(2\frac{\partial Q}{\partial X}(\mu_X,
        \mu_Y)\frac{\partial Q}{\partial Y}(\mu_X, \mu_Y)\right) Cov[X, Y] \\
       &= \left(\frac{\partial Q}{\partial X}(\mu_X, \mu_Y)\right)^2 \sigma_X^2
        + \left(\frac{\partial Q}{\partial Y}(\mu_X, \mu_Y)\right)^2 \sigma_Y^2 
        +\left(2\frac{\partial Q}{\partial X}(\mu_X,
        \mu_Y)\frac{\partial Q}{\partial Y}(\mu_X, \mu_Y)\right) Cov[X, Y] \\
        \\
        &= \left[\left(\frac{\partial Q}{\partial X}(\mu_X, \mu_Y)\right)^2 
        + \left(\frac{\partial Q}{\partial Y}(\mu_X, \mu_Y)\right)^2\right] \sigma_X^2
        + \left(2\frac{\partial Q}{\partial X}(\mu_X,
        \mu_Y)\frac{\partial Q}{\partial Y}(\mu_X, \mu_Y)\right) Cov[X, Y] \\
\end{align*} 

To evaluate the derivatives, we have:
       
\begin{align}
   \frac{\partial \eta}{\partial X}(\mu_X, \mu_Y) &= \frac{\partial
(X/Y)}{\partial X}(\mu_X, \mu_Y) = \left.\frac{1}{Y}\right|_{(X,Y)=(\mu_X, \mu_Y)} \nonumber\nonumber\\
   &= \frac{1}{\mu_Y} = \frac{1}{N(1-p)} = \nonumber\nonumber\\
   &=\frac{1}{Nq}
   \label{E_dQdX}
\end{align} 

and

\begin{align}
   \frac{\partial \eta}{\partial Y}(\mu_X, \mu_Y) &=
\frac{\partial X/Y}{\partial Y}(\mu_X, \mu_Y) = 
\left.-\frac{X}{Y^2}\right|_{(X,Y)=(\mu_X, \mu_Y)} \nonumber\\
   &= -\frac{\mu_X}{\mu_Y^2} =
   -\frac{Np}{N^2(1-p)^2} \nonumber\\ &=-\frac{p}{Nq^2}
   \label{dQdY}
\end{align}

The covariance between X and Y is given by:

\begin{align}
   \mathrm{Cov}(X, Y)   &= \frac{1}{N} \sum_{i=1}^N (x_i-E(X))(y_i-E(Y))  \nonumber\\
   &= \frac{1}{N} \sum_{i=1}^N (x_i-\mu_X)\Big((N-x_i)-(N-\mu_X)\Big)  \nonumber\\ 
   &= \frac{1}{N} \sum_{i=1}^N (x_i-\mu_X)(-x_i+\mu_X)  \nonumber\\ 
   &= - \frac{1}{N} \sum_{i=1}^N (x_i-\mu_X)^2  \nonumber\\ 
   &= - \sigma_X^2
   \label{XY_covariance}
\end{align}

Then, using the expressions \ref{E_dQdX}, \ref{dQdY} y \ref{XY_covariance}

\begin{align*}
\mathrm{Var}(\eta) 
   &= \left[(\frac{1}{Nq})^2 + (\frac{p}{Nq^2})^2\right] Npq +
       2 \frac{1}{Nq} \frac{p}{Nq^2} Npq \\
       &= \frac{1}{N}\, \left(
       \left(\frac{1}{q^2}+\frac{p^2}{q^4}\right)pq +
       2\frac{p^2}{q^2}  \right) \\
       &= \frac{1}{N}\, \left[ \frac{p}{q} + 2\left(\frac{p}{q}\right)^2 + 
       \left(\frac{p}{q}\right)^3 \right]
\end{align*}

Evaluating in $p=1/\sqrt{2}$ we have that

$$
\frac{p}{q} = \frac{1/\sqrt{2}}{1-1/\sqrt{2}} = \sqrt{2}+1
$$

Then, 

\begin{align}
 && \mathrm{Var}(\eta) \approx \frac{28.14214}{N} 
\label{var_eta_teo}
\end{align}


It can be seen in Figure~\ref{eta_variance} that the theoretical variance of the
expression \ref{var_eta_teo} and values of the variance calculated from 10 or 100
samples of values of $\eta$, obtained in equivalent simulated samples of $\beta$ with
the method of the binomial distribution.

\end{document}